\documentstyle[12pt]{article}
\begin{document}
\title{{\bf INFORMATION LOSS IN BLACK HOLES
AND/OR CONSCIOUS BEINGS?}
\thanks{Alberta-Thy-36-94, hep-th/9411193, to be published
in {\em Heat Kernel Techniques and Quantum Gravity}, edited by S. A.
Fulling (Discourses in Mathematics and Its Applications, No. 4, Texas
A\&M University Department of Mathematics, College Station, Texas,
1995).}}
\author{
Don N. Page
\thanks{Internet address:
don@phys.ualberta.ca}
\\
CIAR Cosmology Program, Institute for Theoretical Physics\\
Department of Physics, University of Alberta\\
Edmonton, Alberta, Canada T6G 2J1
}
\date{(1994 Nov. 25)}

\maketitle
\large
\begin{abstract}
\baselineskip 14.7 pt

     In 1976 Stephen Hawking proposed that information
may be lost from our universe as a pure quantum state
collapses gravitationally into a black hole, which then
evaporates completely into a mixed state of thermal radiation.
Although this proposal is controversial, it is tempting to consider
analogous processes that might occur in certain theories
of consciousness.  For example, one might postulate that
independent degrees of freedom be ascribed to the mental
world to help explain the feeling of a correlation between
one's desires and one's choice of actions.  If so, one might
ask whether information in the physical world can be lost
to such postulated degrees of freedom in the mental world.
Or, one might hypothesize that the mental world can
affect the physical world by modifying the quantum action
for the physical world in a coordinate-invariant way
(analogous to the alpha parameters in wormhole theory).
\\
\\
\end{abstract}
\normalsize
\newpage
\baselineskip 14.7 pt

	As a result of his 1974 calculation \cite{Haw74,Haw75}
of thermal emission from black holes, Hawking argued \cite{Haw76}
that a pure quantum state that underwent
gravitational collapse into a black hole
would end up as a mixed quantum state of thermal radiation.
This proposal has continued to be controversial,
and in the conference lecture being reported here,
I discussed various alternatives and arguments for and
against them.  However, as I have little to add in print
to these arguments since my recent review of this subject
\cite{BHI} and subsequent research paper \cite{IIBHR},
here I shall instead briefly turn to some analogous ideas
that I have been considering concerning consciousness.

	I have developed a framework, which I call
Sensible Quantum Mechanics, for relating conscious
perceptions to the quantum state of the universe and for
interpreting quantum mechanics.  Since the basics of this
framework are also to be published elsewhere \cite{PDM,SQM},
here I shall merely summarize it briefly and then consider
some speculations that go beyond it in a way that is somewhat
analogous to the suggestion of information loss in black holes.

	Sensible Quantum Mechanics is given by the following
two basic postulates:

	{\bf Measure Axiom for Perceptions}:  There is a fundamental
measure $\mu(S)$ for each subset $S$ of the ``mental world,''
the set $M$ of all perceptions $p$.

	{\bf Sensible Quantum Axiom}:  The measure $\mu(S)$
is given by the expectation value of an ``awareness operator''
$A(S)$, a positive-valued-operator (POV) measure \cite{Dav},
in the quantum state of the ``physical world'':
	\begin{equation}
	\mu(S) = \langle A(S) \rangle = \langle\psi|A(S)|\psi\rangle
			= Tr[A(S)\rho],
	\label{eq:1}
	\end{equation}

	Here the third expression applies if the quantum state is
represented by the wavefunction or pure state $|\psi\rangle$, and the
fourth expression applies if the quantum state is represented by the
statistical operator or density matrix $\rho$.  (The second
expression can apply in more general situations, such as in
$C^*$-algebra.)

	Since all sets $S$ of perceptions with $\mu(S) > 0$ really
occur in this framework, it is completely deterministic if the
quantum state and the $A(S)$ are determined:  there are no random or
truly probabilistic elements in this framework of Sensible Quantum
Mechanics.  Nevertheless, because the framework has measures for sets
of perceptions, one can readily use them to calculate quantities that
can be interpreted as conditional probabilities.  One can consider
sets of perceptions $S_1$, $S_2$, etc., defined in terms of
properties of the perceptions.  For example, $S_1$ might be the set
of perceptions in which there is a feeling that the universe is
approximately described by a Friedman-Robertson-Walker model, and
$S_2$ might be the set of perceptions in which there is a feeling
that the age of the universe (at the perceived time) is between ten
and twenty billion years.  Then one can interpret
	\begin{equation}
	P(S_2|S_1)\equiv \mu(S_1\cap S_2)/\mu(S_1)
	\label{eq:2}
	\end{equation}
as the conditional probability that the perception is in the set
$S_2$, given that it is in the set $S_1$.  In our example, this would
be the conditional probability that a perception including the
feeling that the universe is approximately described by a
Friedman-Robertson-Walker model, also has the feeling that at the
time of the perception the age is between ten and twenty billion
years.

	Thus in Sensible Quantum Mechanics, probabilities don't
(apply to) ``matter''; they are only in the ``mind.''

	In order to get quantities associated with a single
perception $p$ and to test and compare different theories, assume the
set $M$ of perceptions is a suitable topological space with a prior
measure
	\begin{equation}
	\mu_0(S) = \int_S d\mu_0(p).
	\label{eq:3}
	\end{equation}
Let
	\begin{equation}
	A(S) = \int_S E(p)d\mu_0(p),
	\label{eq:4}
	\end{equation}
	\begin{equation}
	\mu(S) =  \langle A(S) \rangle = \int_S d\mu(p)
			= \int_S m(p) d\mu_0(p),
	\label{eq:5}
	\end{equation}
	\begin{equation}
	m(p) = \langle E(p) \rangle
		= \langle\psi|E(p)|\psi\rangle = Tr[E(p)\rho],
	\label{eq:6}
	\end{equation}

	A rather natural hypothesis to add to the basic framework of
Sensible Quantum Mechanics is that each $E(p)$, which I call a
perception operator, is a projection operator $P(p)$ onto an
`eigenstate of perception' (perhaps a technically `unphysical' state,
one that does not obey the constraint equations, such as the
Wheeler-DeWitt equations, since presumably $P(p)$ would not commute
with the constraints, but this makes no difference for calculating
its expectation value so long as one has an inner product on the full
space of unconstrained states).  However, one can think of many other
possible restrictions on the form of $E(p)$ \cite{PDM,SQM}.

	Some possibilities for the prior measure $\mu_0(S)$ are the
following:
	\begin{equation}
	\mu_0^{(1)}(S) = N(S)
				= \mbox{number of perceptions in $S$
(if discrete)},
	\label{eq:7}
	\end{equation}
	\begin{equation}
	\mu_0^{(2)}(S) = Tr[A(S)]\;\;
				\mbox{(if awareness operators are of
trace class)},
	\label{eq:8}
	\end{equation}
	\begin{equation}
	\mu_0^{(3)}(S) = \int_S g^{1/2}(p) d^np\;\;
				\mbox{(if $M$ has a Riemannian
metric, e.g.},
	\label{eq:9}
	\end{equation}
	\begin{equation}
	g_{ij}dp^idp^j =
Tr\{[E(p^i+dp^i)-E(p^i)][E(p^j+dp^j)-E(p^j)]\},\;\;
				\mbox{if finite)}.
	\label{eq:10}
	\end{equation}

	Now agreement with observation can be tested by the
typicality of one's perception $p$:  Let $S_{\leq}(p)$ be the set of
perceptions $p'$ with $m(p') \leq m(p)$.  Then one can, in a rather
{\it ad hoc} way, define a `typicality'
	\begin{equation}
	T(p) = \mu(S_{\leq}(p))/\mu(M),
	\label{eq:11}
	\end{equation}
which, in the case in which $m(p)$ varies continuously, has a uniform
distribution between 0 and 1 for $p$ chosen randomly with the
infinitesimal measure $d\mu(p)$.  Therefore, using this particular
criterion and assuming that one's perception $p$ is indeed typical in
this regard, one might say that agreement with observation requires
that the prediction, by the theory in question, of $T(p)$ for one's
observation $p$ be not too much smaller than unity.

	One could of course instead use any other property of
perceptions which places them into an ordered set to define a
corresponding `typicality.'  For example, I might be tempted to order
them according to their complexity, if that could be well defined.
Thinking about this alternative `typicality' leaves me surprised that
my own present perception seems to be highly complicated but
apparently not infinitely so.  What simple complete theory could make
a typical perception have a high but not infinite complexity?

	However, the `typicality' defined by Eq.~(\ref{eq:11}) has
the merit of being defined purely from the prior and fundamental
measures, with no added concepts such as complexity that would need
to be defined.  Furthermore, if one makes one of the proposals of
Eq.~(\ref{eq:7})-(\ref{eq:10}), then both the fundamental and the
prior measures are determined by the quantum state and by either the
awareness operators $A(S)$ or by the perception operators $E(p)$, so
the number of basic elements is rather minimal.

	Returning to the definition given by Eq.~(\ref{eq:10}) for
the typicality, one can use a Bayesian approach and assign prior
probabilities $P(H_i)$ to hypotheses $H_i$ that give particular
theoretical predictions for a measure density $m_i(p)$ over all
perceptions and hence assign a particular typicality $T_i(p)$ to
one's perception.  (For example, one might choose to set
$P(H_i)=2^{-n_i}$, where $n_i$ is the rank of $H_i$ in order of
increasing complexity, in order to give the simpler theories higher
prior probabilities.  To do this, one would need to assume some
particular background knowledge whith respect to which one might
define `complexity.')  Although a perception $p$ can be said to exist
in each theory that gives $m_i(p)>0$, so that the {\it existence} of
the perception has unit probability in each, one can instead take the
{\it typicality} of the perception and interpret its probability of
being so small, which is $T_i(p)$ itself (since the typicality has a
uniform distribution), as the probability associated with the
perception $p$ in the hypothesis $H_i$.  This may then be called the
likelihood of $H_i$ given $p$.  By Bayes' rule, the posterior
conditional probability that one should then rationally assign to the
hypothesis $H_i$, if one followed this prescription of interpreting
the typicality as the conditional probability (given the hypothesis
$H_i$) for one's particular perception $p$, would be
	\begin{equation}
	P(H_i|p)=\frac{P(p|H_i)T_i(p)}{\sum_{j}^{}{P(p|H_j)T_j(p)}}.
	\label{eq:12}
	\end{equation}

	There is the potential problem that the right side of
Eq.~(\ref{eq:11}) may have both numerator and denominator infinite,
which makes the typicality $T_i(p)$ inherently ambiguous.  Then one
might use instead
	\begin{equation}
	T_i(p;S) = \mu_i(S_{\leq}(p)\cap S)/\mu_i(S)
	\label{eq:13}
	\end{equation}
for some set of perceptions $S$ containing $p$ that has $\mu_i(S)$
finite for each hypothesis $H_i$.  This is a practical limitation
anyway, since one could presumably only hope to be able to compare
the measure densities $m(p)$ for some small set of perceptions rather
similar to one's own.  Unfortunately, this makes the resulting
$P(H_i|p)$ depend on this chosen $S$ as well as on the other
postulated structures.

	In Sensible Quantum Mechanics, the measure for all sets $S$
of perceptions in the mental world is determined by the physical
world (i.e., the quantum state), along with the awareness operators.
This seems to leave mysterious the correlation between will and
action:  why do I do as I please?  That is, why is my desire to do
something I feel am capable of doing apparently correlated with my
perception of actually doing it?  One might seek to explain this by
going beyond Sensible Quantum Mechanics to a framework, say to be
called Sensational Quantum Mechanics, in which the mental world acts
back on the physical world.

	One way in which this could be done, going beyond the
framework of Sensible Quantum Mechanics without violating it, would
be for desires in the mental world to affect the action functional
that is used in a particular Feynman path integral to define the
quantum state.  If this modification of the action were done in a
coordinate-invariant way, it would not violate local energy-momentum
conservation.  Instead, this would be analogous to the way in which
Coleman \cite{Col} proposed that baby universes may adjust effective
coupling constants in our universe (such as setting the cosmological
constant or gravitating energy density to zero) via wormholes (the
analogue of Descartes' pineal glands \cite{Des}).

	There are also other less conservative ways in which one
might postulate that mind affects matter.  For example, one might
imagine a modification of the action in a coordinate-dependent way,
so that coordinate invariance is broken.  Or, one might imagine that
the quantum state is modified in such a way that it cannot be given
by any Feynman path integral with any form of the action.

	Another way in which the mental world might affect the
physical world is by changing the boundary conditions of the path
integral that defines the quantum state in terms of the actional
functional.  As in all the ways mentioned so far, once the modified
quantum state of the physical world is determined, the framework of
Sensible Quantum Mechanics could apply without alteration.

	One might go on to consider mind-body interactions which
violate the framework of Sensible Quantum Mechanics.  For example,
one might consider quantum theories in which the wavefunction really
collapses.  Then one might imagine that the mental world could
conceivably influence the time at which a collapse occurs, the
eigenbasis in which the collapse takes place, and the specific
eigenvector outcome (if it indeed results in a pure state).

	Another proposal that could not be incorporated within
Sensible Quantum Mechanics is that the measure on sets of conscious
perceptions is not determined by expectation values in the quantum
state but rather is determined by the trajectory in configuration
space that is an essential part of the formalism of Bohm's version of
quantum mechanics \cite{Boh}.  This by itself need not lead to any
influence of mind on matter, but one could imagine modifications in
which the mental world affects the Bohmian trajectories, say by
providing a new consciousness force.

	Finally, to get to a proposal in which one can consider the
possibility of information loss in conscious beings, one might
hypothesize that conscious perceptions are new perceptual degrees of
freedom that simply have not been included previously in physics.
For example, they might be quantum variables, new arguments of the
wavefunction of the extended world.  (Alternatively, they might be
semiclassical variables in a hybrid semiclassical theory, so that the
perceptions themselves, and not just the measures on sets of them,
are determined by expectation values of the quantum physical
variables and may also act back to affect the quantum state itself.
However, I shall not consider this possibility further here.)

	Presumably these new quantum variables of conscious
perceptions would interact with the old ones (which I shall continue
to call the physical world or matter, since it or its quantum state
is what we are now familiar with as the physical world, though I
would not object to the terminology of someone else who might prefer
to combine both sets of quantum variables in his definition of the
physical world, so long as we are clear in each discussion which
terminology is being used).  This interaction represents the effect
of matter on mind and of mind on matter.

	Such an interaction allows the conceptual possibility,
analogous to what Hawking proposed for black holes, of information
loss in conscious beings.  For example, one could imagine an initial
state which is a product state of a pure state for the (old) physical
variables and some arbitrary state of the mental variables.  But then
as the two variables interact, one gets an entangled state that is no
longer a product state and hence no longer gives a pure state for the
physical variables when traced out over the mental variables.

	So far, this possibility is analogous to information loss
into a persisting black hole, which is not very controversial (though
the allowed amount might be, particularly for small persisting black
hole remnants).  But one could ask the question of what happens if
one starts with no conscious beings, they evolve out of a state that
was initially purely physical, and then they die out.  Can the
excitations of the mental variables disappear with the conscious
beings, say into a mental singularity, taking the information they
had with them, so that in the end it no longer resides within the
physical universe?  This would be analogous to what Hawking proposes
happens to the information that falls into a black hole (that had
formed out of an initial non-black-hole state) when (and if) the
black hole completely disappears.  (Of course, this information loss
in conscious beings is by no means required even if conscious beings
appear and disappear within the system, since they could be like
photons that are emitted and absorbed without any loss of unitarity
in ordinary systems \cite{Gol}.  However, here I am exploring the
conceptual possibility that consciousness {\it might} lead to a loss
of information as Hawking proposes black holes do.  Admittedly, the
analogy between consciousness and black holes is tenuous.  The main
similarly is that we don't yet understand the quantum nature of
either.)

	To test whether information is lost in conscious beings, one
apparently would need to construct a closed physical system in a
particular pure state, wait a sufficient time for conscious beings to
evolve and die out, and then measure the system.  With a sufficiently
large ensemble of similarly-prepared systems and a sufficiently large
set of different measurements on the final systems (a number at least
of the order of the square of the dimension of the relevant Hilbert
space for each system), one could in principle determine the density
matrix of each and thereby determine to some approximation whether it
was pure (assuming the systems were sufficiently isolated that they
remained uncorrelated, and assuming that they each evolved the same
way to identical final density matrices).  Of course, in practice it
would be impossible to perform most of the large number of
measurements needed, and there might even be restrictions in
principle to performing them, since measurements are restricted to
the interactions actually occurring in nature, and since one has the
restrictions of the properties of the quantum state of the universe
(e.g., the second law of thermodynamics) \cite{Gol}.

	Even if the measurements were really possible in principle,
which I shall assume for the sake of argument, the experiment would
be extremely difficult, since one might need to wait for a few
billion years for conscious beings to evolve and die out.  One would
presumably also need each system to be large in order to contain an
energy source such as the sun to last long enough in it.  There also
needs to be a large enough garbage dump for the waste energy.

	For example, in shining at its present luminosity for the age
of the earth, the sun produces about $5.5\times 10^{50}$ ergs of
energy.  If this is to be uniformly distributed as thermal radiation
over a volume large enough that the temperature be low enough for the
thermal disequilibrium necessary for life on earth, say below the
freezing point of water, one needs a volume bigger than about
$1.3\times 10^{55} \rm{cm^3}$, which fills out a sphere of radius
larger than about half a parsec or 1.5 light years.  In units of
Boltzmann's constant, the entropy of this radiation is more than
about $1.94\times 10^{64}$.  The exponential of this gives a rough
estimate of the dimension of the relevant Hilbert space, so one might
need far more than $10^{10^{64.59}}$ measurements to determine
approximately all the relevant elements of the final density matrix
of each system (presuming each is identical and uncorrelated).

	One might well need much larger individual systems and
corresponding many, many more measurements if one wanted a
significant probability that conscious beings would evolve in each.
There is of course the staggering problem of where to put all of
these enormously many systems, since they must be kept far enough
apart that they do not interact enough to become sufficiently
correlated.  (A crude estimate of the requirement that the tidal
interaction energy between the systems, varying as the inverse cube
of the distance, be less than the splitting of the energy levels
within one of the systems, puts the minimum distance between them as
more than $10^{10^{63.81}}$ light years  The difference between the
upper exponent here and that previous is very nearly the common
logarithm of twice three, or six, and different plausible assumptions
about the power law in the distance-dependence of the interaction
energy would simply change this integer a bit but would leave the
minimum separation distance utterly enormous.)

	Thus the experimental problem of determining whether or not
information is fundamentally lost in conscious beings seems utterly
impractical.  However, it is still trivial compared to the problem of
determining experimentally whether or not information is lost down
black holes of the solar mass $M_{\odot}=9.14\times 10^{37}$ in
Planck units, which have thermal entropies around $4\pi
M^2_{\odot}\sim 10^{76.66}$ and would require more than
$10^{10^{76.96}}$ measurements to give a rough determination of the
final density matrix after black hole evaporates.

	One can also amusingly calculate that if information is not
lost, the Poincar\'{e} recurrence time of an isolated black hole in a
rigid nonpermeable box with stationary boundary conditions should be
of the order of the exponential of the number of energy eigenstates
with significant quantum components, which itself should be of the
order of the exponential of the thermal entropy.  Therefore, for a
black hole of the solar mass, the Poincar\'{e} recurrence time should
be of the order of
	\begin{equation}
	t_{\mbox{\scriptsize{Poincar\'{e}}}} \sim \exp{\exp{(4\pi
M^2_{\odot})}}
					\sim 10^{10^{10^{76.66}}}\;\;
					\mbox{Planck times, millenia,
or whatever}.
	\label{eq:14}
	\end{equation}
For a black hole containing the mass within the presently visible
region of our universe, it should be of the order of
	\begin{equation}
	t_{\mbox{\scriptsize{Poincar\'{e}}}} \sim
10^{10^{10^{10^{2.08}}}}\;\;
					\mbox{Planck times, millenia,
or whatever}.
	\label{eq:15}
	\end{equation}
Finally, if one takes the mass within what may be the entire universe
in one of Linde's stochastic inflationary models \cite{Lin} with a
massive inflaton whose mass is about $m=10^{-6}$ in Planck units, and
puts this mass into a black hole in a suitable box, one should get a
Poincar\'{e} recurrence time of the order of
	\begin{equation}
	t_{\mbox{\scriptsize{Poincar\'{e}}}} \sim
\exp{\exp{\exp{(4\pi m^{-2})}}}
					\sim
10^{10^{10^{10^{10^{1.1}}}}}\;\;
					\mbox{Planck times, millenia,
or whatever}.
	\label{eq:16}
	\end{equation}
So far as I know, these are the longest finite times that have so far
been explicitly calculated by any physicist.

	I should again emphasize that I do not have strong reasons
for proposing that information is actually lost in conscious beings,
but it does seem to be a logical possibility, on a rather similar
level to Hawking's proposal \cite{Haw76} of information loss down
black holes.  However, it is obvious from my papers \cite{BHI,IIBHR}
that I am sceptical of the latter, and I am similarly sceptical of
the former.

	On the other hand, even if information is not fundamentally
lost down either black holes or conscious beings, so that the
evolution that includes them in the intermediate states is described
by a unitary $S$-matrix, one can still ask whether these $S$-matrices
are predictable in terms of presently-known physics.  In the black
hole case it is fairly obvious that the $S$-matrix is not, because we
do not yet have a complete understandable quantum theory of gravity.
In the case of conscious beings, if their evolution can occur in
situations in which nonlinear quantum gravity and black holes have
negligible effects or probabilities of occurring, one might suppose
that the $S$-matrix should be accurately described by the present
Standard Model of particle physics (presumably augmented by enough
gravitational interactions to give a good approximation to Solar
System dynamics).

	Of course, we do not know the parameters of the Standard
Model nearly accurately enough for a calculation of the evolution of
any large piece of matter for any significant time (even for a
calculation in principle, by computational or even noncomputational
means with unlimited resources).  But one might ask whether there is
{\it any} precise choice of the unknown parameters, in the present
form of the Standard Model Lagrangian (including a sufficient amount
of gravity), that would give the $S$-matrix up to the accuracy that
one would expect from, say, its neglect of quantum nonlinear gravity
and of Grand Unified processes such as baryon decay.

	To test this by the procedure outlined above of actually
measuring an ensemble of initial and final systems seems inordinately
difficult.  One might hope to be able to test it for a single
conscious being, but to do that without letting the being evolve
naturally in a very large system over a very long time, one would
need to know how to create it, which might be difficult.  Although by
measuring the physical states of enough conscious beings, one might
learn how to create them on scales much smaller than the Solar
System, the scale and thermal entropy would still presumably be
sufficiently large that the number of measurements needed would still
be utterly enormous.  (For example, a human of mass 70 kg has a
dimensionless thermal entropy on the order of $2\times 10^{28}$ and
so would require of the order of $10^{10^{28.24}}$ measurements to
determine the $S$-matrix.)

	If these measurements could be done and were consistent with
unitary evolution (so that no information is fundamentally {\it lost}
in the conscious beings), then I would not be surprised to find that
the $S$-matrix is not consistent with the Standard Model with any
choice of its parameters.  Whether the difference would be attributed
to other new physics that has no particular connection with
consciousness, or whether it would be attributed to effects of
consciousness, is an interesting question that will have to wait
until we gain a better understanding of these matters.

\section*{Acknowledgments}

\hspace{.25in}I appreciate discussions I have had on this general
subject with those listed in \cite{BHI,PDM}, and particularly recent
discussions with Shelly Goldstein (whom I especially thank for
perceptive comments on the original version of this paper),
Tom\'{a}\v{s} Kopf, Werner Israel, Michael Lockwood, and Euan
Squires.  Finally, financial support has been provided by the Natural
Sciences and Engineering Council of Canada.

\end{document}